\documentclass[12pt,preprintnumbers,amsmath,amssymb,prd,floatfix,superscriptaddress,nofootinbib]{revtex4}
\usepackage{graphicx}
\usepackage{epsfig}
\usepackage{bm}
\usepackage{amsfonts}
\usepackage{epstopdf}
\begin{document}

\title{\bf Dynamics of Particles Around a Schwarzschild-like Black Hole
in the Presence of Quintessence and Magnetic Field }

\author{Mubasher Jamil}
\email{mjamil@sns.nust.edu.pk;
jamil.camp@gmail.com}\affiliation{School of Natural Sciences (SNS),
National University of Sciences and Technology (NUST), H-12,
Islamabad, Pakistan}

\author{ Saqib Hussain }
\affiliation{School of Natural
Sciences (SNS), National University of Sciences and Technology
(NUST), H-12, Islamabad, Pakistan}

\author{ Bushra Majeed }
\affiliation{School of Natural Sciences (SNS), National University
of Sciences and Technology (NUST), H-12, Islamabad, Pakistan}

\begin{abstract}
{\bf Abstract:} We investigate the dynamics of a neutral and a
charged particle around a static and spherically symmetric black
hole in the presence of quintessence
 matter and external
magnetic field. We explore the conditions under which the particle
moving around the black hole could escape to infinity after
colliding with another particle. The  innermost stable circular
orbit (ISCO) for the particles are studied in detail. Mainly the
dependence of ISCO on dark energy and on the presence of external
magnetic field in the vicinity of black hole is discussed. By using
the Lyapunov exponent, we compare the stabilities of the orbits of
the particles in the presence and absence of dark energy and
magnetic field. The expressions for the center of mass energies of
the colliding particles near the horizon of the black hole are
derived. The effective force on the particles due to dark energy and
magnetic field in the vicinity of black hole is also discussed.
\end{abstract}
 \maketitle

\newpage
\section{Introduction}
The accelerating expansion of the universe indicates the presence of
elusive dark energy. The presence of dark energy is supported by
several astrophysical observations including the study of Ia
Supernova \cite{Ia1}, cosmic microwave background (CMB) \cite{cmb}
and large scale structure (LSS) \cite{lss,lss1}. The nature of dark
energy is not understood until now. It is explained by cosmological
models in which dominant factor of dark energy density may possess
negative pressure such as cosmological constant $\Lambda$ with a
state parameter $w_{q}=-1$. There are  other scalar field models
that are proposed  such as  quintessence \cite{q}, phantom dark
energy \cite{jamil}, k-essence \cite{k}, holographic dark
energy\cite{hdee} to name a few. Dark energy is approximately $70$
percent of the energy density of the universe. If dark energy is
dynamical, then naturally it will become more dominant in the future
and will play a crucial role at all length scales. In this context,
we study the motion of the particles around the black hole
surrounded by dark energy and magnetic field.

\par
The magnetic coupling  process is responsible for  attraction of black hole
with its accretion disc \cite{3}. According to this process, the angular momentum
and energy are transferred from a black hole to its surrounding disc.
There exists a sufficient strong magnetic field in the vicinity of
black hole (Penrose effect for magnetic field) \cite{4,5}.
Observational evidences indicate that magnetic field should be
present in the vicinity of  black holes \cite{new}. This magnetic
field arises due to  plasma in the surrounding of black hole. The
relativistic motion of particles in the conducting matter in the
accretion disk may generate the magnetic field inside the disk. This
field does not affect the geometry of the black hole, yet it affects
the motion of  charged particles and support them to escape
\cite{1,2}.
\par
Ba\~{n}ados, Silk and West (BSW) proposed that some black holes may
act as particle accelerators \cite{bsw}. In the vicinity of extremal
Kerr black hole, they have found that infinite center-of-mass energy
(CME) can be achieved during the collision of particles. The BSW
effect has been studied for different black hole spacetimes
\cite{bsw}-\cite{ms1}. In this paper we obtain the CME expression
for the colliding particles near horizon of Schwarzschild-like black
hole surrounded by quintessence matter and BSW effect is studied for
both neutral and charged particles.
\par
 Quintessence is defined as
a scalar field coupled to gravity with the potential which decreases
as field increases \cite{q3}. The solution for a spherically
symmetric black hole surrounded by quintessence matter was derived
by Kiselev \cite{Kiselev}. It has the state parameter in the range ,
$-1<w_{q}<\frac{-1}{3}$. In this work we will focus on Kiselev
solution.
 Null geodesics around Kiselev black hole have been studied in \cite{q2}. We consider the Kiselev solution in the presence of external axi-symmetric magnetic field, which is homogeneous at infinity. This magnetic field and
quintessence matter strongly affect the dynamics of the particles
and location of their innermost stable circular orbits (ISCO) around
black hole. Before dealing with dynamics of a
charged particle around black hole, we do analysis of a neutral
particle without considering magnetic field. We construct the
dynamical equations from the Lagrangian formalism which  are not
solvable via analytic methods. Even in the absence of black hole,
charged particle motion in the non-uniform magnetic field which is chaotic \cite{8,9}.
\par
 Main objective of our work is to study the circumstances under which the initially moving particle would escape from
ISCO or its motion would remain bounded, after collision with some other particle. The effect of collision on the energy of
the particle is also discussed. We have calculated
the velocity of a particle required to escape to infinity and
investigate some characteristics of the particle's motion moving
around black hole. With the help of Lyapunov exponent, a comparison of the stability of orbits for
massive and massless particles is also established \cite{Lyapunov1}.
\par
 The outline of the paper is as follows:  We
explain our model in section II and derive an expression for escape
velocity of the neutral particle. In section III dynamics of a charged particle is discussed, we derive the
equations of motion. In section IV dimensionless form of the equations are given. In section V CME
expressions are derived for colliding particles. In section VI the
Lyapunov exponent is calculated. In section VII the force on the
charged particle is calculated. We discuss the trajectories for escape energy and
escape velocity of the particle in section VIII.
Concluding remarks are given in section IX. We use $(+,-,-,-)$ sign
convention and gravitational units, $c=1$, in this work.

\section{Dynamics of a Neutral Particle}
The geometry of
static spherically symmetric black hole surrounded by the
quintessence matter (Kiselev solution) is given by \cite{Kiselev}
\begin{eqnarray}\label{1}
ds^{2}&=&f(r)dt^{2}-\frac{1}{f(r)}dr^{2}-r^{2}d\theta^{2}-r^{2}\sin^{2}\theta
d\phi^{2},
       \nonumber\\&&
       f(r)=1-\frac{2M}{r}-\frac{c}{r^{3w_{q}+1}}.
\end{eqnarray}
Here $M$ is the mass of black hole, $c$ is the quintessence
parameter and $w_{q}$ has range $-1<w_{q}<\frac{-1}{3}$ while we
will focus on $w_{q}=\frac{-2}{3}$. Metric $(\ref{1})$
has curvature singularity at $r=0$. For $f(r)=0$ we
get two values of $r$:
\begin{equation}
  r_{+}=\frac{1+\sqrt{1-8Mc}}{2c}, ~~~r_{-}=\frac{1-\sqrt{1-8Mc}}{2c}.
\end{equation}
The region $r=r_{-}$ corresponds to black hole horizon while
$r=r_{+}$ represents the cosmological horizon.  Therefore, $r_{-}$
and $r_{+}$ are the two coordinate singularities in the metric
$(\ref{1})$. If $8Mc=1$, we get the degenerate solution for the
spacetime at $r_{\pm}=\frac{1}{2c}$ and if $8Mc>1$, horizons do
not exist. For very small value of $c$, $r_{+}\approx\frac{1}{c}$.
Furthermore, we can say that the restriction on $c$, is
$c\leq\frac{1}{8M}$.
\par
We  discuss the dynamics of a neutral particle in the
Schwarzschild-like background defined by $(\ref{1})$. There are
three constants of motion corresponding $(\ref{1})$ in which two of
them arise as a result of two Killing vectors \cite{13}
\begin{equation}\label{1b}
\xi_{(t)}=\xi^{\mu}_{(t)}=\partial_{t} , \qquad
\xi_{(\phi)}=\xi^{\mu}_{(\phi)}=\partial_{\phi}.
\end{equation}
where $\xi^{\mu}_{t}=(1,0,0,0)$ and $\xi^{\mu}_{\phi}=(0,0,0,1)$, Eq.
$(\ref{1b})$ implies that, black hole metric $(\ref{1})$ is
invariant under time translation and rotation around symmetry axis
$(\theta=0)$. The corresponding conserved quantities (conjugate
momenta) are the  energy  per unit mass $\mathcal{E}$ and azimuthal
angular momentum  per unit mass, $L_{z}$, respectively given by
\begin{equation}\label{2}
  \mathcal{E}\equiv f(r)\dot{t},
\end{equation}
\begin{equation}\label{3}
  -L_{z}\equiv\dot{\phi}r^{2}\sin^{2}\theta.
\end{equation}
\par
Here over dot represents  differentiation with respect to proper
time $\tau$. The third constant of motion is the total angular
momentum of the particle i.e.
\begin{equation}\label{18}
  L^{2}=(r^2\dot{\theta})^{2}+\frac{L_{z}^{2}}{\sin^{2}\theta}=r^{2}v^{2}_{\bot}+\frac{L_{z}^{2}}{\sin^{2}\theta}.
\end{equation}
Here we denote $v_{\bot}\equiv -r\dot{\theta}_{o}$. By using the
normalization condition of $4$-velocity $u^{\mu}u_{\mu}=1$ and
constants of motion $(\ref{2})$ and $(\ref{3})$, we get the equation
of motion of neutral particle
\begin{equation}\label{4}
\dot{r}^{2}=\mathcal{E}^{2}-(1+\frac{L_{z}^2}{r^{2}\sin^{2}\theta})f(r).
\end{equation}
At the turning points of the moving particles from the trajectories
$\dot{r}=0$, hence equation $(\ref{4})$ gives
\begin{equation}\label{17}
  \mathcal{E}^{2}=(1+\frac{L_{z}}{r^{2}\sin^{2}\theta})f(r)\equiv
  U_\text{eff},
\end{equation}
where $U_\text{eff}$ is the effective potential.
\par
Consider a particle in the circular orbit $r=r_{o}$, where $r_{o}$
is the local minima of the effective potential. This orbit exists
for $r_{o}\in(4M,\infty)$. Generally for non-degenerate case
$(r_{+}\neq r_{-})$ the energy and azimuthal angular momentum
corresponding to local minima $r_{o}$ are
\begin{equation}
 L_{zo}= \frac{\sqrt{cr_{o}^{2}-2M}}{\sqrt{c+\frac{6M-2r_{o}}{r_{o}^{2}}}},
\end{equation}
and
\begin{equation}
  \mathcal{E}_{o}=\frac{2\big(2M+r_{o}(cr_{o}-1)\big)^{2}}{r_{o}\big(6M+r_{o}(cr_{o}-2)\big)}.
\end{equation}
For the degenerate case which is defined by $c=\frac{1}{8M}$ or
$r_{+}=r_{-}$. The energy and azimuthal angular momentum
corresponding to $r_{o}$ are
\begin{equation}
 L_{zo}= \frac{\sqrt{\frac{r_{o}^{2}}{8M}-2M}}{\sqrt{\frac{1}{8M}+\frac{6M-2r_{o}}{r_{o}^{2}}}},
\end{equation}
and
\begin{equation}
  \mathcal{E}_{o}=\frac{2\big(2M+r_{o}(\frac{r_{o}}{8M}-1)\big)^{2}}{r_{o}\big(6M+r_{o}(\frac{r_{o}}{8M}-2)\big)}.
\end{equation}

The ISCO is defined by $r_{o}=4M$ which is the convolution point of
the effective potential \cite{14}. We have not restricted ourself to
this local minima at $r_{o}$ because  it depends on the applied
condition which we will discuss later in section \ref{7a}.

\par
Now consider the particle is in a ISCO and collides with another
particle, the later one is coming from the rest position at infinity
as a freely falling particle. After collision between particles,
three cases are possible for the particles: (i) remain bounded
around black hole, (ii) capture by black hole and (iii) escape to
infinity. The results depend on the collision process. For small
changes in energy and angular momentum, orbit of the particle is
slightly perturbed but particle remains bounded. For larger change
in energy and angular momentum, it can go away from initial path and
could be captured by black hole or escape to infinity.

After the collision particle should have new values of energy and
azimuthal angular momentum and the total angular momentum.
 We simplify the problem by applying the following
conditions: $(i)$ the azimuthal angular momentum does not change and
$(ii)$ initial radial velocity remains same after collision. Under
these conditions only energy can change by which we can determine
the motion of the particle. After collision particle acquires an
escape velocity $(v_{\perp})$ in orthogonal direction of the
equatorial plane \cite{new}.

After collision the total angular momentum and energy of the
particle  become (at $\theta=\frac{\pi}{2}$)
\begin{equation}
  L^{2}=r_{o}^{2}v_{\perp}^{2}+L_{z}^{2},
\end{equation}
\begin{equation}\label{33}
 \mathcal{E}=\bigg[f(r)\bigg(1+\frac{(L_{z}+rv_{\perp})^2}{r^{2}}\bigg)\bigg]^{\frac{1}{2}}.
\end{equation}
\par
These new values of angular momentum and energy are greater from
their values
 before collision because during collision colliding particle may
impart some of its energy to the orbiting particle. We get the
expression $(\ref{33a})$ for velocity $v$ from Eq. $(\ref{33})$
after solve it for $v$ to get
\begin{equation}\label{33a}
  v_{\perp}^\text{esc}\geq\frac{L_{z}r(r-2M-cr^{2})+\sqrt{r^{4}(r(1-cr)-2M)(2M+r(cr+\mathcal{E}^{2}-1))}}{r^{2}(2M+r(cr-1))},
\end{equation}
particle would escape if $| v_{\perp}^\text{esc}|\geq v_{\perp}$.

\section{Dynamics of a Charged Particle}

We investigate how does the motion of a charged particle is effected
by both magnetic field in the black hole exterior and gravitational
field. The general  Killing vector equation is \cite{10}
\begin{equation}\label{34}
  \Box\xi^{\mu}=0,
\end{equation}
where $\xi^{\mu}$ is a Killing vector. Eq. (\ref{34}) coincides with
the Maxwell equation for 4-potential $A^{\mu}$ in the Lorentz gauge
$A^{\mu}_{\ \ ;\mu}=0$. The special choice \cite{6}
\begin{equation}
  A^{\mu}=\frac{\mathcal{B}}{2}\xi^{\mu}_{(\phi)},
\end{equation} corresponds to the test magnetic field,
where $\mathcal{B}$ is the magnetic field strength. The
$4$-potential is invariant under the symmetries which corresponds to
the Killing vectors as discussed above, i. e.,
\begin{equation}
  L_{\xi}A_{\mu}=A_{\mu,\nu}\xi^{\nu}+A_{\nu}\xi^{\nu}_{,\mu}=0.
\end{equation}
A magnetic field vector is defined as \cite{13}
\begin{equation}\label{5}
  \mathcal{B}^{\mu}=-\frac{1}{2}e^{\mu\nu\lambda\sigma}F_{\lambda\sigma}u_{\nu},
\end{equation}
where
\begin{equation}\label{6}
  e^{\mu\nu\lambda\sigma}=\frac{\epsilon^{\mu\nu\lambda\sigma}}{\sqrt{-g}},\ \
  \epsilon_{0123}=1,\ \ g=det(g_{\mu\nu}).
\end{equation}\\
$\epsilon^{\mu\nu\lambda\sigma}$ is the Levi Civita symbol. The
Maxwell tensor is defined as
\begin{equation}\label{7}
  F_{\mu\nu}=A_{\nu,\mu}-A_{\mu,\nu}=A_{\nu;\mu}-A_{\mu;\nu}.
\end{equation}
For a local observer at rest in the space-time $(\ref{1})$, the non-vanishing components of $4$- velocity are
\begin{equation}\label{8}
  u^{\mu}_{0}=\frac{1}{\sqrt{f(r)}}\xi^{\mu}_{(t)},~~~~~~~u^{\mu}_{3}=\frac{1}{\sqrt{r^{2}\sin^{2}\theta}}\xi^{\mu}_{(\phi)}.
\end{equation}
The other two components $u^{\mu}_{1}$ and $u^{\mu}_{2}$  are zero
at the turning point $(\dot{r}=0)$. From equations
$(\ref{5})-(\ref{8})$ we have obtained the magnetic field given
below
\begin{equation}\label{19}
  \mathcal{B}^{\mu}=\mathcal{B}\frac{1}{\sqrt{f(r)}}\Big[\cos\theta\delta^{\mu}_{r}-\frac{\sin\theta\delta^{\mu}_{\theta}}{r}\Big].
\end{equation}
Here we considered magnetic field to be directed along the vertical
($z$-axis) and $\mathcal{B}>0$.
\par
 The Lagrangian of the particle of mass $m$ and electric charge $q$ moving
in an external magnetic field of a curved space-time is given by
\cite{19}
\begin{equation}\label{10}
  \mathcal{L}=\frac{1}{2}g_{\mu\nu} u^{\mu}u^{\nu}+\frac{qA_{\mu}}{m}u^{\mu},
\end{equation}
and generalized 4-momentum of the particle $p_{\mu}=m
u_{\mu}+qA_{\mu}$. The new constants of motion are
\begin{equation}\label{11}
  \dot{t}=\frac{\mathcal{E}}{f(r)},\ \ \ \
  \dot{\phi}=\frac{L_{z}}{r^{2}\sin^{2}\theta}-B,
\end{equation}
here
\begin{equation}\label{12}
  B\equiv\frac{q\mathcal{B}}{2m}.
\end{equation}
By using these constants of motion in the Lagrangian we get the
dynamical equations for $\theta$ and $r$ respectively
\begin{eqnarray}\label{13}
  \ddot{\theta}=B^{2}\sin\theta\cos\theta-\frac{2}{r}\dot{r}\dot{\theta}-\frac{L_{z}^{2}\cos^{2}\theta}{r^{4}\sin^{3}\theta},
\end{eqnarray}
\begin{eqnarray}\label{14}
  \ddot{r}&=&-\frac{(2M+r(cr-1))^{2}}{(-2M+r-cr^{2})}\big(\frac{L_{z}^{2}}{r^{4}\sin^{2}\theta}\big)+
  \frac{(2M+cr^{2})}{(2M+cr^{2}-r)^{2}}(\mathcal{E}+\dot{r}^{2})
  \nonumber\\&&
  +\frac{(2M+r(cr-1))^{2}}{(-2M+r-cr^{2})}(B^{2}\sin^{2}\theta+\dot{\theta}^{2}).
\end{eqnarray}
By using normalization condition we get
\begin{eqnarray}\label{15}
  \mathcal{E}^{2}=\dot{r}^{2}+r^{2}f(r)\dot{\theta}^{2}+f(r)\big[1+r^{2}\sin^{2}\theta\big(\frac{L_{z}}{r^{2}\sin^{2}\theta}-B\big)^{2}\big].
\end{eqnarray}
From Eq. (\ref{15}) we can write the effective potential as
\begin{equation}\label{40}
  U_{eff}=f(r)\big[1+r^{2}\sin^{2}\theta\big(\frac{L_{z}}{r^{2}\sin^{2}\theta}-B\big)^{2}\big].
\end{equation}
The above equation is a constraint i.e. if it is satisfied initially,
then it is always valid, provided that  $\theta(\tau)$ and $r(\tau)$
are controlled by equation $(\ref{13})$ and $(\ref{14})$.\\
Let us discuss the symmetries of Eqs. $(\ref{10})-(\ref{15})$, these
equation are invariant under the transformation given below
\begin{equation}\label{16}
  \phi\rightarrow-\phi,\ \ L_{z}\rightarrow-L_{z},\ \ B\rightarrow-B.
\end{equation}
Therefore, without losing the generality, we consider the positively
charged particle. The  trajectory of a negatively charged particle is
related to positive charge's trajectory by transformation
$(\ref{16})$. If we make a choice $\mathcal{B}>0$  then we will have
to study both cases when $L_{z}>0,\ L_{z}<0$. They are physically
different: the change of sign of $L_{z}$ means the change of
direction of the Lorentz force on the particle.

System of Eqs. $(\ref{10})-(\ref{15})$ is invariant with respect to
reflection $(\theta\rightarrow\pi-\theta)$. This transformation
retains the initial position of the particle and changes
$v_{\perp}\rightarrow-v_{\perp}$. Therefore, it is sufficient to
consider only the positive value of $v_{\perp}$.

\section{Dimensionless Form of the Dynamical Equations}
\par
\begin{figure}[!ht]
\centering
\includegraphics[width=9cm]{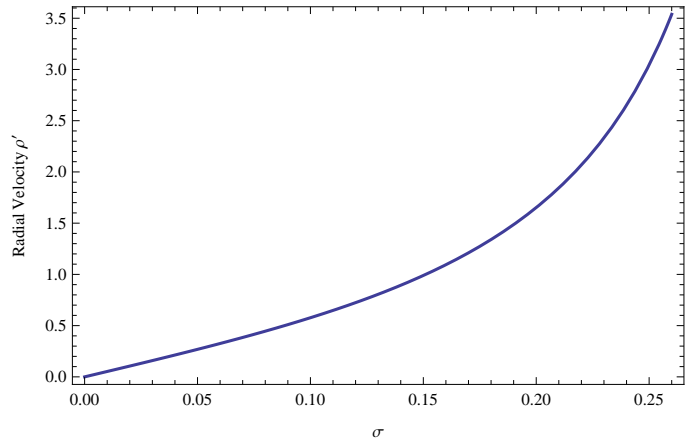}
\caption{Radial velocity as a
function of $\sigma$ for $\ell=5$, $\mathcal{E}=1$, $b=0.25$ and
$c_{1}=0.125$.}\label{f1.13}
\end{figure}

Before integrating our dynamical equations of $r$ and $\theta$
numerically we make these equations dimensionless by introducing
following dimensionless quantities \cite{11,13}:
\begin{equation}
  2m=r_{d} ~~~~~~ \sigma=\frac{\tau}{r_{d}},\ \rho=\frac{r}{r_{d}},\ \ell=\frac{L_{z}}{r_{d}},\ b=Br_{d},~~~~ c_{1}=c r_{d}.
\end{equation}
Eqs. $(\ref{13})$ and $(\ref{14})$ acquire the form
\begin{eqnarray}\label{1.11}
  \frac{d^{2}\theta}{d\sigma^{2}}=b^{2}\sin\theta\cos\theta-\frac{2}{\rho}\frac{d\rho}{d\sigma}\frac{d\theta}{d\sigma}+
  \frac{\ell^{2}\cos^{2}\theta}{\rho^{4}\sin^{3}\theta},
\end{eqnarray}
\begin{eqnarray}\label{1.12}
  \frac{d^{2}\rho}{d\sigma^{2}}&=&(1+c_{1}\rho^{2}-\rho)\frac{\ell^{2}}{\rho^{4}\sin^{2}\theta}
          +\frac{(1+c_{1}\rho^{2})}{(1+c_{1}\rho^{2}-\rho)^{2}}\bigg(\mathcal{E}+\big(\frac{d\rho}{d\sigma}\big)^{2}\bigg)
          \nonumber\\&&
          -(1+c_{1}\rho^{2}-\rho)\bigg(b^{2}\sin^{2}\theta+\big(\frac{d\theta}{d\sigma}\big)^{2}\bigg).
\end{eqnarray}
These are very complicated equations and to study such problems
$3$D numerical simulation of the magnetohydrodynamics (MHD) in a strong gravitational field is required \cite{16}.
For simplicity we fix $\theta=\frac{\pi}{2}$ then the Eq. $(\ref{1.11})$ is
satisfied and the Eq. $(\ref{1.12})$ becomes
 \begin{eqnarray}\label{1.13}
  \frac{d^{2}\rho}{d\sigma^{2}}&=&(1+c_{1}\rho^{2}-\rho)\bigg(\frac{\ell^{2}}{\rho^{4}}-b^{2}\bigg)
          +\frac{(1+c_{1}\rho^{2})}{(1+c_{1}\rho^{2}-\rho)^{2}}\bigg(\mathcal{E}+\big(\frac{d\rho}{d\sigma}\big)^{2}\bigg).
\end{eqnarray}
We have solved Eq. $(\ref{1.13})$ numerically by using the built
in command NDSolve in Mathematica $8.0$. We have obtained the
interpolating function as a solution of Eq. $(\ref{1.13})$ and
plotted the derivative of interpolating function (radial velocity of
the particle) as a function of $\sigma$ in Fig. $\ref{f1.13}$, it shows that $\rho^{'}(\sigma)$ is increasing as we increase the dimensionless parameter
$\sigma$.

Eqs. $(\ref{15})$ and (\ref{40}) become
\begin{eqnarray}
  \mathcal{E}^{2}=\big(\frac{d\rho}{d\sigma}\big)^{2}+\rho^{2}\big(1-c_{1}\rho-\frac{1}{\rho}\big)\big(\frac{d\theta}{d\sigma}\big)^{2}+U_\text{eff}.
\end{eqnarray}
\begin{equation}
  U_\text{eff}=\big(1-c_{1}\rho-\frac{1}{\rho}\big)\bigg[1+\rho^{2}\big(\frac{\ell}{\rho^{2}\sin^{2}\theta}-b\big)^{2}\bigg].
\end{equation}
The energy of the particle moving around the black hole in an orbit
of radius $\rho_{o}$ at the equatorial plane is given by
\begin{equation}\label{21}
  U_\text{eff}=\mathcal{E}_{o}^{2}=\big(1-c_{1}\rho-\frac{1}{\rho}\big)\bigg[1+\frac{(\ell-b\rho_{o}^{2})^{2}}{\rho_{o}^{2}}\bigg].
\end{equation}
Solving $\frac{dU_\text{eff}}{d\rho}=0$ and
$\frac{d^{2}U_\text{eff}}{d\rho^{2}}=0$ simultaneously, we calculate
$b$ and $\ell$ in term of $\rho$,
\begin{equation}\label{1.6}
  \frac{dU_\text{eff}}{d\rho}=\rho^{2}-2b\ell\rho^{2}-b^{2}\rho^{4}-2b^{2}\rho^{5}(c_{1}-1)+\ell^{2}(3+2\rho(c_{1}-1)),
\end{equation}and
\begin{equation}\label{1.7}
  \frac{d^{2}U_\text{eff}}{d\rho^{2}}=2\big[2b\ell\rho^{2}-\rho^{2}+b^{2}\rho^{5}(1-c_{1})-3\ell^{2}(2+\rho(c_{1}-1))\big].
\end{equation}
The obtained expressions for $b$ and $\ell$ are
\begin{eqnarray}\label{1.8}
  b&=&\frac{1}{2\sqrt{2\rho^{4}(1+\rho(c_{1}\rho-1))^{2}\big[3+\rho\big(\rho(4+c_{1}(18+\rho(3c_{1}\rho-18)))-10\big)\big]}}
  \nonumber\\&&
  \times\bigg[\sqrt{\rho^{4}\big[1-\rho(3+c_{1}\rho(\rho(3c_{1}\rho-1)-6))\big]\big(\rho(1+c_{1}\rho(6+\rho(c_{1}\rho-3)))-3\big)^{3}}
  \nonumber\\&&
  +\rho^{2}\big(\rho(1+c_{1}\rho(6+\rho(c_{1}\rho-3)))-3\big)(3+\rho(\rho(4+c_{1}(14+3\rho(c_{1}\rho-3)))-9))\bigg]^{\frac{1}{2}},
\end{eqnarray}
and \begin{eqnarray}\label{1.9}
  \ell&=&\frac{\sqrt{\rho^{4}\big[1-\rho(3+c_{1}\rho(\rho(3c_{1}\rho-1)-6))\big]\big(\rho(1+c_{1}\rho(6+\rho(c_{1}\rho-3)))-3\big)^{3}}}
  {\big(\rho(1+c_{1}\rho(6+\rho(c_{1}\rho-3)))-3\big)^{2}}
  \nonumber\\&&
  \times\bigg[\frac{1}{2\sqrt{2\rho^{4}(1+\rho(c_{1}\rho-1))^{2}\big[3+\rho\big(\rho(4+c_{1}(18+\rho(3c_{1}\rho-18)))-10\big)\big]}}
  \nonumber\\&&
  \times\bigg\{\sqrt{\rho^{4}\big[1-\rho(3+c_{1}\rho(\rho(3c_{1}\rho-1)-6))\big]\big(\rho(1+c_{1}\rho(6+\rho(c_{1}\rho-3)))-3\big)^{3}}
  \nonumber\\&&
  +\rho^{2}\big(\rho(1+c_{1}\rho(6+\rho(c_{1}\rho-3)))-3\big)(3+\rho(\rho(4+c_{1}(14+3\rho(c_{1}\rho-3)))-9))\bigg\}^{\frac{1}{2}}\bigg].
\end{eqnarray}
\begin{figure}[!ht]
\centering
\includegraphics[width=11cm]{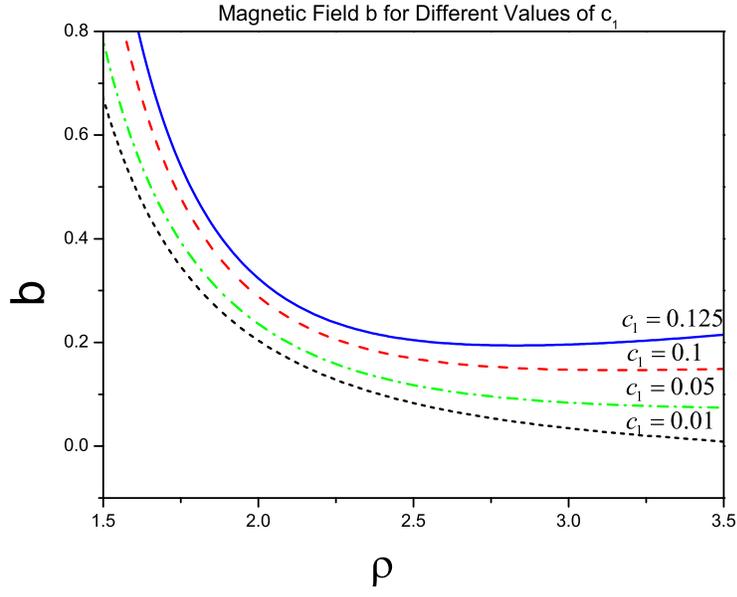}
\caption{Magnetic field $b$ as a
function of $\rho$ for different value of $c_{1}$.} \label{f1.11}
\end{figure}
\begin{figure}[!ht]
\centering
\includegraphics[width=11cm]{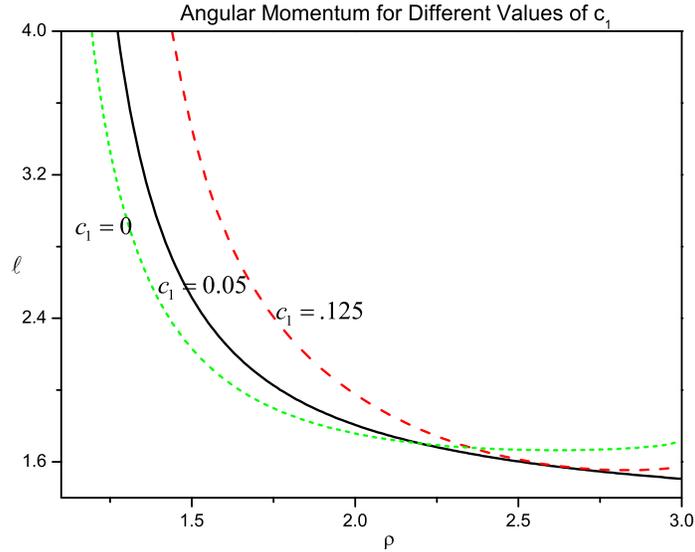}
\caption{Figure shows the behavior of angular momentum as a
function of $\rho$ for different value of $c_{1}$.} \label{f1.12}
\end{figure}
\begin{figure}[!ht]
\centering
\includegraphics[width=11cm]{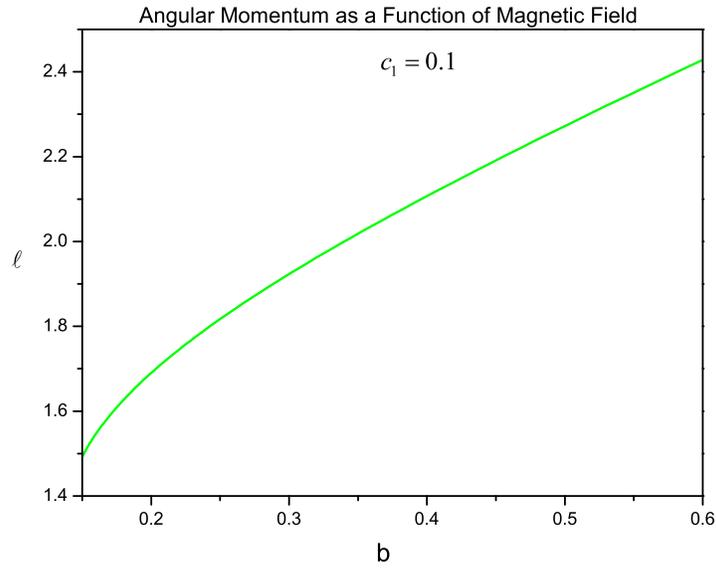}
\caption{Behavior of angular momentum
$\ell_{+}$ vs magnetic field $b$.} \label{b12}
\end{figure}
\begin{figure}[!ht]
\centering
\includegraphics[width=11cm]{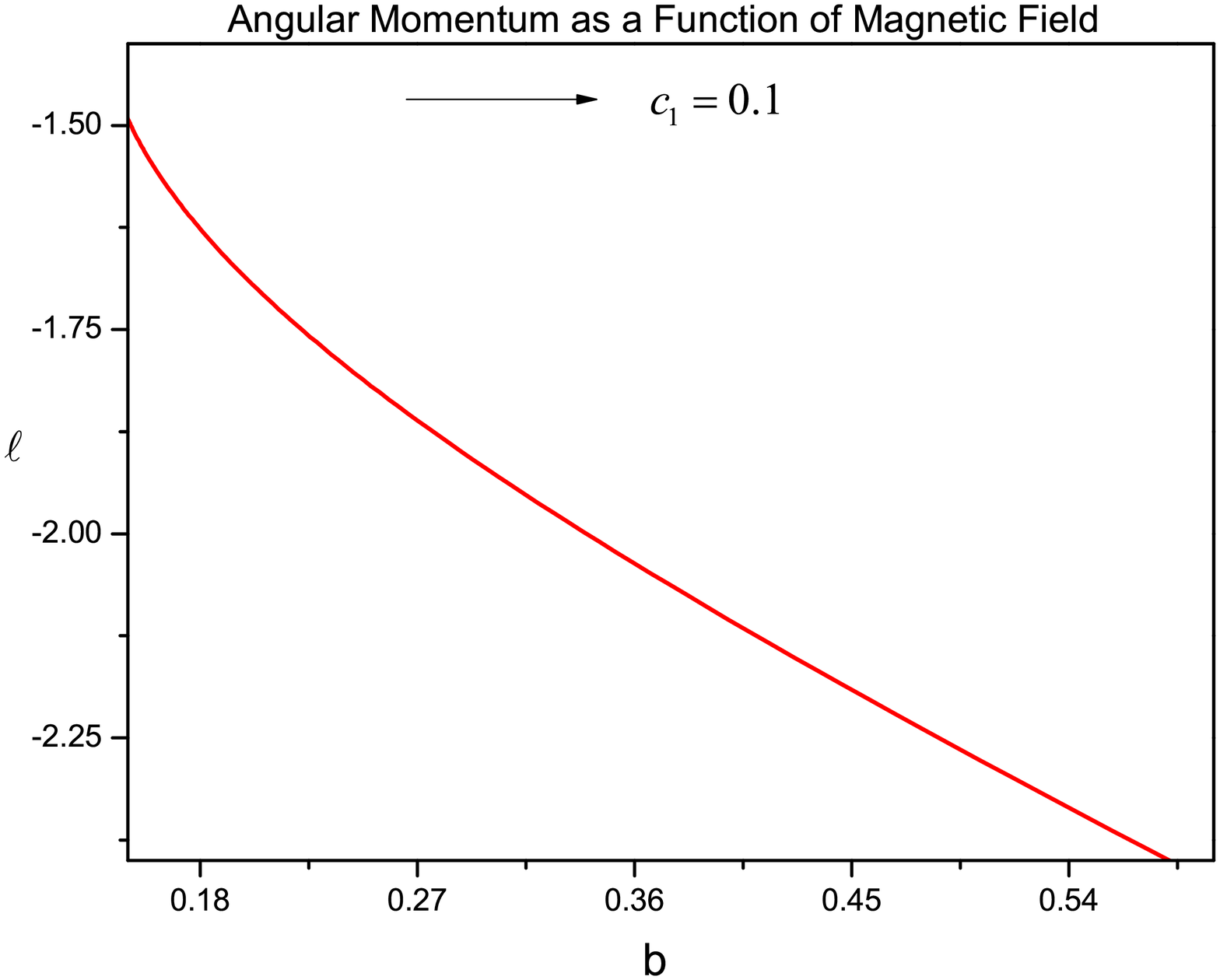}
\caption{Angular momentum
$\ell_{-}$ against magnetic field $b$.} \label{a12}
\end{figure}
In Fig. \ref{f1.11} we have plotted magnetic field $b$ against
$\rho$ for different values of $c_{1}$. It can be seen that the
strength of magnetic field is increasing for large value of $c_{1}$.
We can conclude that the presence of dark energy strengthens the
magnetic field present in the vicinity of black hole. The
strength of magnetic field  is decreasing away from the black hole.
In Fig. \ref{1.12} we have plotted the $\ell$ for different value of $c_{1}$. It can be
seen that larger the value of $c_{1}$ greater will be $\ell$.
So, presence of dark energy increases the Lorentz force on
the orbiting particle.
The angular momentum $\ell$, for
ISCO, as function of magnetic field $b$ is shown in Fig. \ref{b12} and Fig. \ref{a12}. Lorentz force is attractive if $\ell>0$, corresponds to Fig. \ref{b12} and it is repulsive if
$\ell<0$, corresponds to Fig. \ref{a12}.
\par
As we did before in the case of a neutral particle, we assume that
the collision does not change the azimuthal angular momentum of the
particle but it does change the  velocity $v_{\perp}>0$. Due to this,
the particle energy changes, $\mathcal{E}_{o}\rightarrow\mathcal{E}$,
and is given by
\begin{equation}\label{20}
  \mathcal{E}=\sqrt{\bigg(\big(1-c_{1}\rho-\frac{1}{\rho}\big)\bigg[1+\rho^{2}\big(\frac{\ell+\rho v_{\perp}}{\rho^{2}}-b\big)^{2}\bigg]\bigg)}.
\end{equation}
Escape velocity of the particle obtained from Eq. (\ref{20}) is given below
\begin{eqnarray}\label{31}
  v_{\perp}^\text{esc}&\geq&\frac{1}{\rho^{2}(1+\rho(c_{1}\rho-1)}
  \big[\rho(1+\rho(c_{1}\rho-1))(b\rho^{2}-\ell)
  \nonumber\\&&
  \pm\sqrt{\rho^{4}(\rho(1-c_{1}\rho)-1)(1+\rho(\mathcal{E}^{2}+c_{1}\rho-1))}\big].
\end{eqnarray}
Behavior of escape velocity is discussed graphically in section VIII.
\section{Center of Mass Energy of the Colliding Particles}
\subsection{In the absence of magnetic field ($\textbf{B=0}$)}
First we consider that the two neutral particles of masses $m_1$ and
$m_2$ coming from infinity collide near the black hole when there is
no magnetic field. The collision energy of the particles of masses
$m_1=m_2=m_o$ in the center of mass frame is defined as \cite{bsw}
\begin{equation}\label{ecm} E_{cm}=m_o\sqrt{2}\sqrt{1-g_{\mu\nu} u_1^{\mu}
u_2^{\nu}},\end{equation}where  \begin{equation} u^{\mu}_i\equiv
\frac{dx^{\mu}}{d\tau},~~~i=1, ~2
\end{equation}
is the $4$-velocity of each of the particles. Using Eqs. (\ref{2}),
(\ref{3}), and (\ref{4}), in Eq. (\ref{ecm}) we get the CME for the
neutral particle, falling freely from rest at infinity, given below
\begin{eqnarray}\label{aa9}
E_{cm}&=& m_o\sqrt{2} \Big[2+(\frac{L_1^2 + L_2^2}{2
r^2})(\frac{\mathcal{E}^2 + f(r)}{\mathcal{E}^2})+ (L_1
L_2)\Big(\frac{f(r)(L_1 L_2)- 2 r^2 \mathcal{E}^2}{2 \mathcal{E}^2
r^4}\Big)+ \frac{f(r)}{2 \mathcal{E}^2} \Big]^{1/2}.
\end{eqnarray}
 We are interested to find out the CME of the particles near the
 horizon, so taking $f(r)=0$ we get
\begin{equation}\label{a10}
E_{cm}=2m_o\sqrt{1+\frac{1}{4r_h^2}(L_1- L_2)^2},
\end{equation}where $r_h= \frac{1\pm \sqrt{1-8 M c}}{2c}$ represents the horizons of the black hole, obtained
earlier. The expression of CME obtained in Eq. (\ref{a10}) could be
infinite if the angular momentum of one of the particles gets
infinite value, but it would not allow the particle to reach the
horizon of the black hole. Thus the CME in Eq. (\ref{a10}) can not
be unlimited.

\subsection{In the presence of magnetic field}
For a charged particle moving around the black hole we have obtained
the constants of motion defined in Eq. (\ref{11}), using it with
normalization condition of the metric we get equation of motion
of the particle
\begin{equation} \dot{r}^2= \mathcal{E}^2- f(r)[1+
r^2(\frac{L}{r^2}-B)^2]\label{rdot}.\end{equation} Using Eqs.
(\ref{11}), (\ref{rdot}) with Eq. ({\ref{ecm}}) we get the
expression for CME of the charged particles coming from infinity,
colliding near the black hole
\begin{eqnarray}\label{ecm1}
E_{cm}&=& m_o\sqrt{2} \Big[2+ (L_1^2+ L_2^2)\Big[\frac{f(r)+
\mathcal{E}^2}{2 r^2 \mathcal{E}^2}+\frac{f(r) B^2}{2
\mathcal{E}^2}\Big]- (L_1+ L_2)\Big[B(\frac{\mathcal{E}^2+
f(r)}{\mathcal{E}^2})+ \frac{f(r) r^2B^3}{\mathcal{E}^2}\Big]+
\nonumber \\&& L_1 L_2\Big[\frac{f(r)}{2 \mathcal{E}^2 r^4}\Big(L_1
L_2+2Br^2(L_1+ L_2)+4 B^2 r^4\Big)-\frac{1}{r^2}\Big]+\nonumber\\&&
B^2 r^2(\frac{ \mathcal{E}^2+f(r)}{ \mathcal{E}^2})+\frac{f(r)}{2
\mathcal{E}^2}+ \frac{f(r) r^4 B^4}{2
 \mathcal{E}^2}\Big]^{1/2},
\end{eqnarray}
near horizon i.e. at $f(r)=0$, Eq. (\ref{ecm1}) becomes
\begin{equation}\label{aa10}
E_{cm}=2m_o\Big[1+\frac{1}{4 r_h^2}(L_1-L_2)^2- \frac{B}{2}[(L_1+
L_2)- Br^2]\Big]^{1/2},
\end{equation}where $r_h$ represent the horizons of the black hole.
The CME in Eq. (\ref{aa10}) could be infinite, if the angular
momentum of one of the particles has infinite value, for which the
particle can not reach the horizon of the black hole. Thus the CME
defined in Eq. (\ref{aa10}) is some finite energy.

\section{Lyapunov Exponent for the instability of orbit}
\begin{figure}[!ht]
\centering
\includegraphics[width=8.5cm]{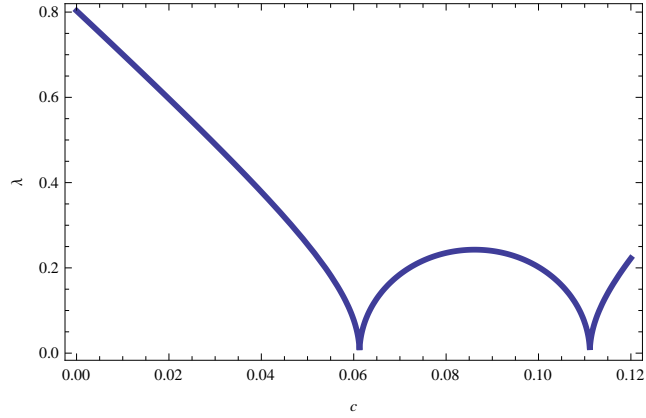}
\caption{Lyapunov exponent as a function of $c$ for
massive particle. Here $r=3$ $M=1$, $L=3.22$, and $b=0.25$}
\label{f12}
\end{figure}

\begin{figure}[!ht]
\centering
\includegraphics[width=8.5cm]{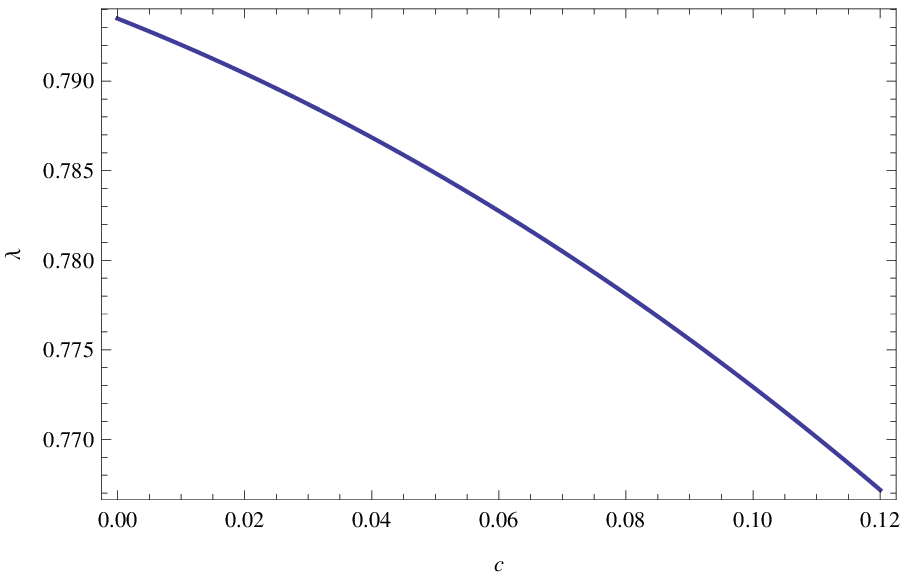}
\caption{Lyapunov exponent as a function of $c$ for
massless particles. Here $r=3$, $M=1$, $L=3.22$, and $b=0.25$ } \label{f12a}
\end{figure}

\begin{figure}[!ht]
\centering
\includegraphics[width=11cm]{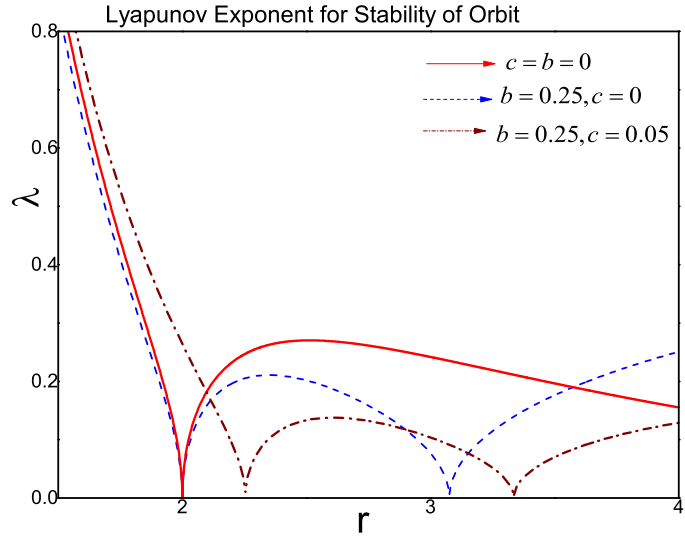}
\caption{Lyapunov exponent for different values of $c$ and $b$ as a
function of radial coordinate $r$ for
 $M=1$, and $L=3.22$.} \label{13a}
\end{figure}

We can check the instability of circular orbit by Lyapunov exponent, $\lambda$,
which is given by \cite{Lyapunov1}
\begin{equation}\label{1.1}
\lambda=\sqrt{\frac{-{U}_{eff}^{\prime\prime}(r_{o})}{2\dot{t}(r_{o})^{2}}}
\end{equation}
\begin{equation}\label{1.2}
  \lambda=\sqrt{\frac{(2M+r(-1+cr))\big(-2Mr^{2}+4BLMr^{2}+B^{2}r^{5}(1-3cr)-L^{2}(12M+r(cr-3))\big)}{L^{2}r^{4}}}
\end{equation}
Behavior of $\lambda$ as a function of $c$ is shown in Fig. \ref{f12} and Fig. \ref{f12a}. It can be seen from these figures
 that instability of circular orbits
is less for non zero $c$ compared to the case when $c=0$, i.e. around Schwarzschild black
hole. In Fig. \ref{13a}, $\lambda$ is compared
for three different types of black hole, Schwarzschild black hole,
Schwarzschild black hole immersed in magnetic field and Kiselev black hole immersed in
magnetic field. It is observed that with non zero
$c$ or $B$, stability is more as compared to Schwarzschild black
hole, $\lambda$ is smaller for Kiselev black hole as
compared to schwarzschild black hole and it gets more small for Kiselev black hole surrounded by both dark energy and magnetic field. In all the figures here after, keep in mind that
 $b=B r_{d}$ .
\pagebreak
\section{Effective Force on the Particle}
Effective force on a particle, computed with the help of effective potential is given by
\begin{equation}\label{1.3}
  F=\frac{-1}{2}\frac{dU_\text{eff}}{dr},
\end{equation}
\begin{equation}\label{1.5}
  F=-\frac{M}{2r^{4}}\big(6L^{2}-4BLr^{2}-2B^{2}r^{4}\big)+\frac{1}{2r^{4}}\big(2L^{2}r-2r^{2}-2Br^{5}\big)
       -\frac{c}{2r^{4}}\big(r^{2}L^{2}+2BLr^{4}-r^{4}-3B^{2}r^{6}\big).
\end{equation}
First and third terms of equation $(\ref{1.5})$ are responsible for attractive force if
$\big(6L^{2}>-4BLr^{2}-2B^{2}r^{4}\big)$ and
$\big(r^{2}L^{2}+2BLr^{4}>-r^{4}-3B^{2}r^{6}\big)$ respectively. Here third term  is appearing due to quintessence matter, hence the quintessence is behaving as a source of attraction inside the horizons. Second term is repulsive
if $\big(2L^{2}r>-2r^{2}-2Br^{5}\big)$. In case of
massless particles the first term and the dark energy term are purely responsible for
attractive force contribution (without any condition) and remaining term is responsible for repulsive force contribution \cite{q2}.
\begin{figure}[!ht]
\centering
\includegraphics[width=12cm]{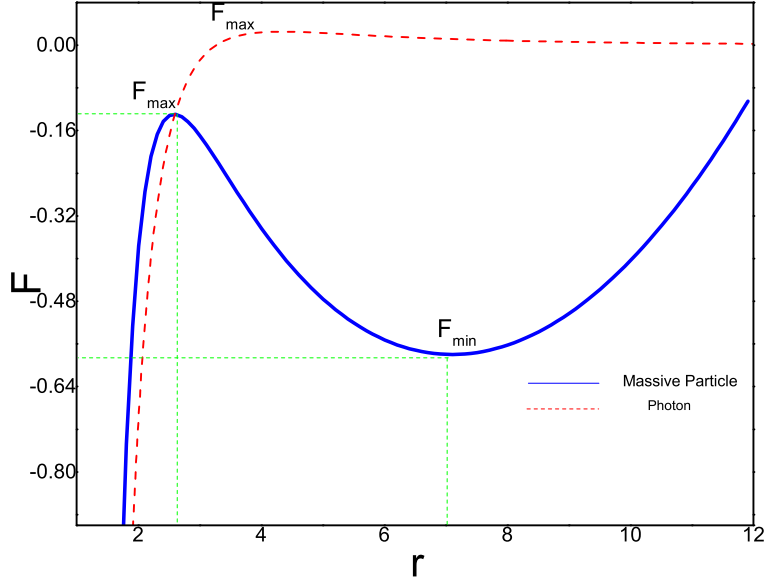}
\caption{Effective force as a function of
$r$. Here $c=0.05$, $M=1$, $L=3.22$ and $b=0.5$.} \label{f13}
\end{figure}
\par
In Fig. \ref{f13} we have plotted the effective force for an
orbiting particle in a circular orbit. In Fig. \ref{f13} $F_{max}$ correspond to
unstable circular orbits and $F_{min}$ corresponds to stable circular orbits.
From this figure we can deduce that there is no stable circular orbits for photon,
it only exist for massive particle.
For the rotational (angular) variable
\begin{equation}\label{25}
\frac{d\phi}{d\tau}=\frac{L_{z}}{r^{2}}-b.
\end{equation}
If right hand side of equation $(\ref{25})$ is positive
$(L_{z}>b)$, then the Lorentz force  on the particle is repulsive
(particle moving in anticlockwise direction), and if right hand side is
negative $(L_{z}<b)$, then the Lorentz force is attractive
(clockwise rotation).
\section{Trajectories for Effective Potential and Escape Velocity}\label{7a}
\begin{figure}[!ht]
\centering
\includegraphics[width=11cm]{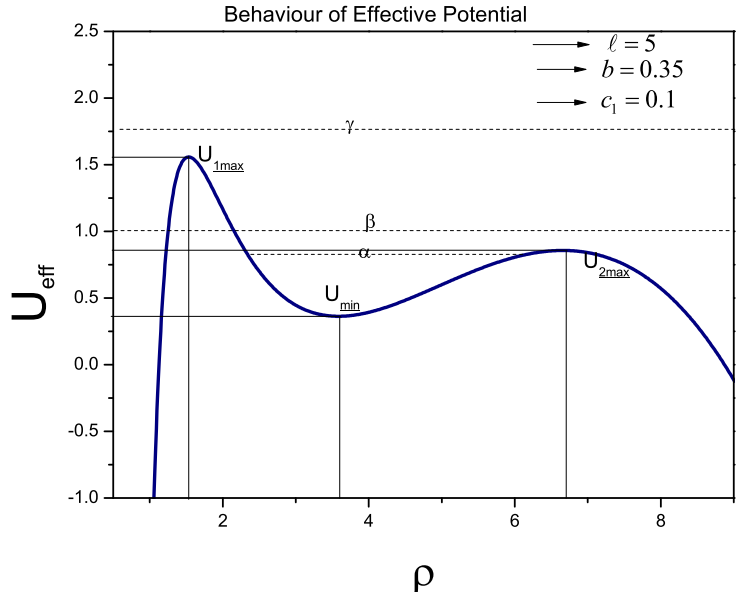}
\caption{The effective potential
versus $\rho$.}\label{f7}
\end{figure}

Behavior of effective potential is demonstrated by plotting it vs $\rho$, in Fig. \ref{f7}. The
horizontal line $\alpha$ with $\mathcal{E}<1$ corresponds to bound
motion, this is the analogue of elliptical motion in Newtonian
theory. The trajectories of the particle is not closed in general.
The line segment $\beta$ with $\mathcal{E}>1$ corresponds to a
particle coming from infinity and then move back to infinity
(hyperbolic motion). 

\begin{figure}[!ht]
\centering
\includegraphics[width=11cm]{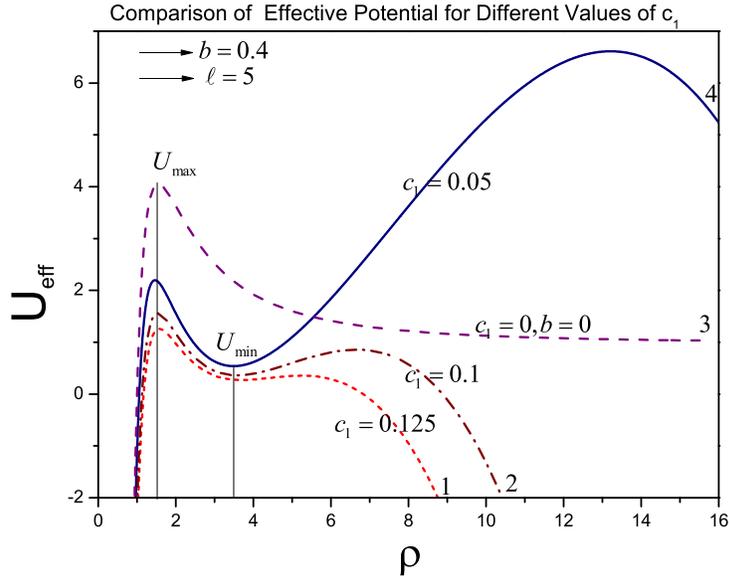}
\caption{The effective potential as a
function $\rho$, for different value of $c$.} \label{f8}
\end{figure}
The line $\gamma$ does not intersects with the
curve of effective potential and passes above its maximum value
$U_{1max}$. It corresponds to particle which  is falling into the
black hole (captured by the black hole) and $U_{1max}$ and $U_{2max}$ correspond to unstable orbits and
$U_{min}$ refers to a stable circular orbit.

\begin{figure}[!ht]
\centering
\includegraphics[width=11cm]{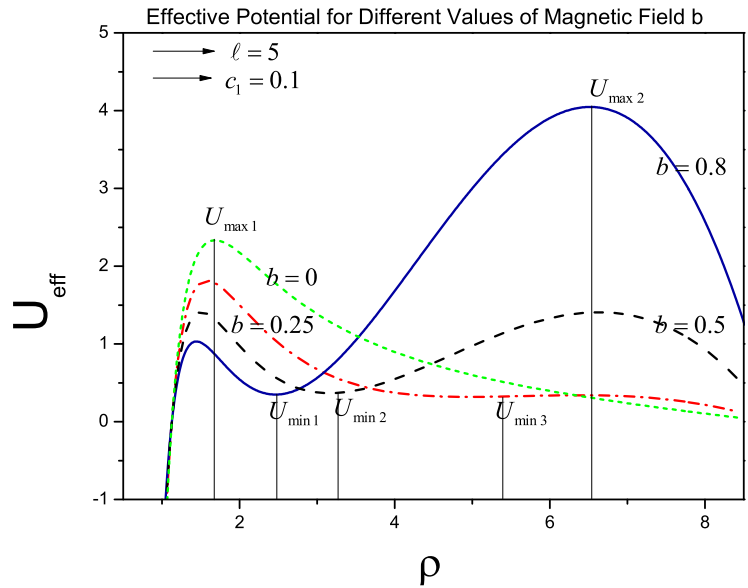}
\caption{The effective potential
against $\rho$, for different value of magnetic field $b$.}
\label{a8}
\end{figure}

\begin{figure}[!ht]
\centering
\includegraphics[width=11cm]{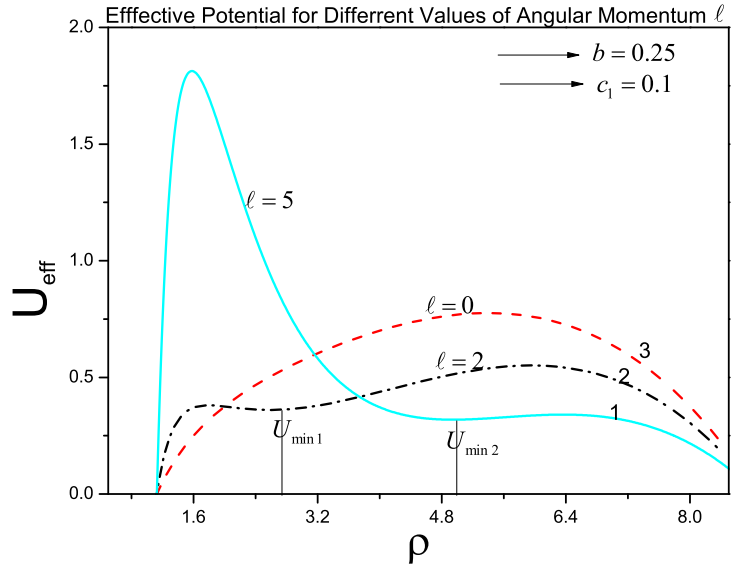}
\caption{The effective potential against
$\rho$, for different value of angular momentum $\ell$.} \label{a9}
\end{figure}

In Fig. \ref{f8} we have compared the effective potentials for
different values of $c$. One can notice as the value of $c$ increases
the maxima and minima of effective potential shifted downward. Here
$U_{max}$ and $U_{min}$ corresponds to unstable and stable  orbits
of the particle around the black hole respectively. In Fig.
\ref{f8}, curve $3$ represents
 the Schwarzschild  effective potential \cite{15}. Therefore, one can say that the dark energy acts to decrease the effective potential. We can conclude that force on the particle  due
to dark energy is attractive. Hence the possibility for a particle
to be captured by the black hole is greater due to presence of dark
energy as compare to the case when $c=0$. Effective potential vs
$\rho$ is plotted in Fig. \ref{a8} for different values of
magnetic field $b$. One can notice from the Fig. \ref{a8} that orbits are more stable in the presence of magnetic field as
compare to the case when magnetic field is absent, $b=0$. It can
also be seen that the local minima of the effective potential which
corresponds to ISCO is shifting toward the horizon which  is in
agreement with \cite{6,17}. We have compared the effective potential
for massive and massless particles (photons) in Fig. \ref{a9}. For photon there
is no stable orbit as there is no minima for $\ell=0$ represented by
plot $3$ in Fig. \ref{a9}. While for the massive particle there
are local minima $U_{min1}$ and $U_{min2}$ which correspond to
stable orbits. It can also be concluded that the particle having
larger value of angular momentum $\ell$ can escape easily as compare
to the particle with lesser value of angular momentum $\ell$.

\begin{figure}[!ht]
\centering
\includegraphics[width=9cm]{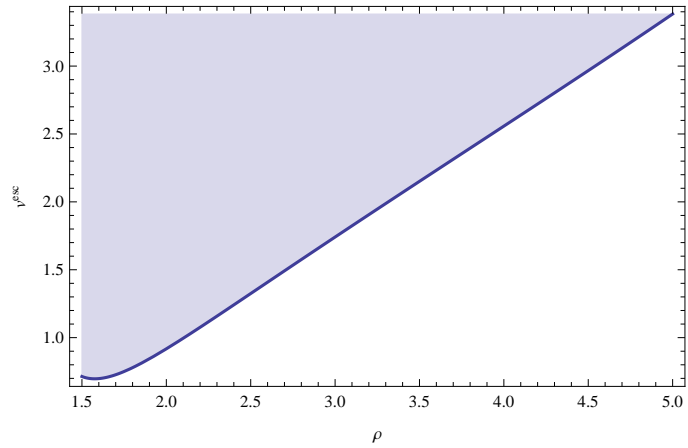}
\caption{The escape velocity as a function of
$\rho$ for $\ell=3.22$, $b=0.50$ and $c_{1}=0.10$.}\label{f16}
\end{figure}

\par
\begin{figure}[!ht]
\centering
\includegraphics[width=11cm]{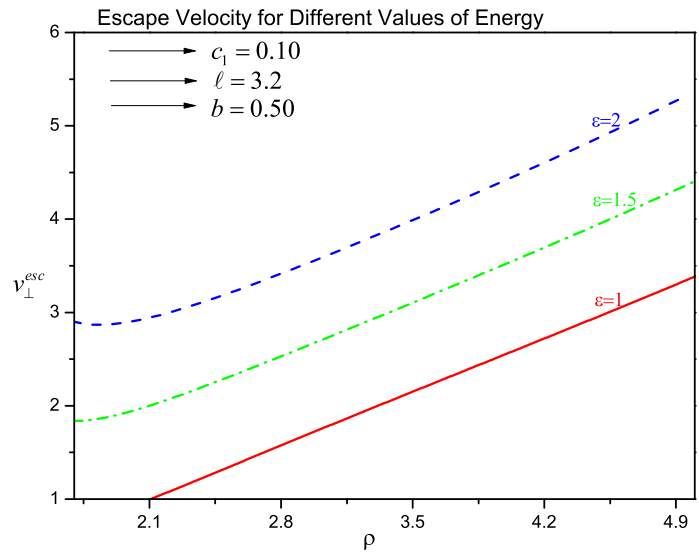}
\caption{Plot of escape velocity $v_{esc}$ as
a function of  $\rho$ for different values of energy
$\mathcal{E}$.}\label{a16}
\end{figure}
\begin{figure}[!ht]
\centering
\includegraphics[width=11cm]{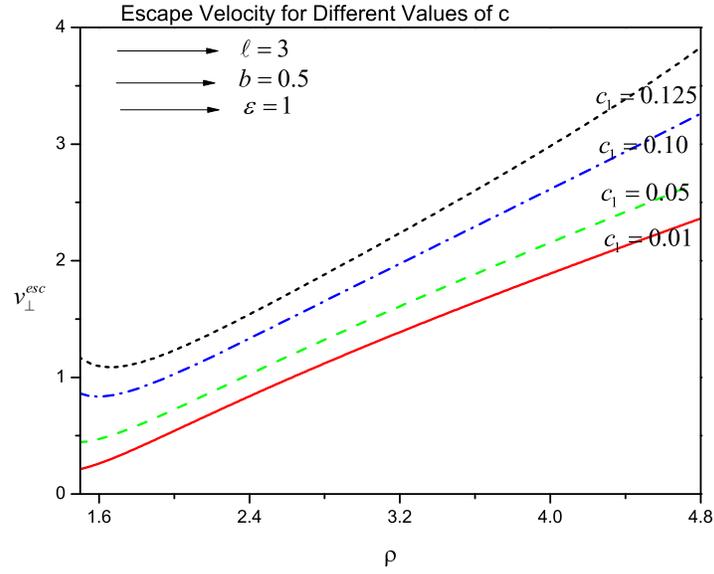}
\caption{Plot of escape velocity $v_{esc}$
against $\rho$ for different values of $c_{1}$.}\label{b16}
\end{figure}
\begin{figure}[!ht]
\centering
\includegraphics[width=11cm]{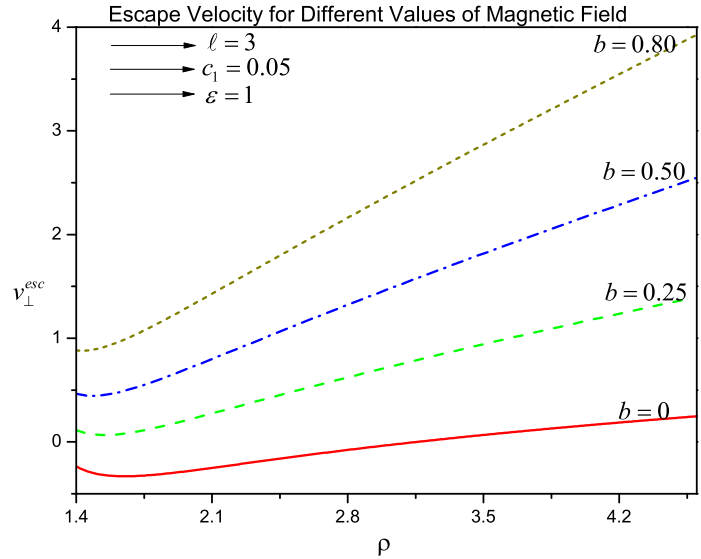}
\caption{Plot of escape velocity $v_{esc}$ vs
$\rho$ for different values of magnetic field $b$.}\label{c16}
\end{figure}

Fig. \ref{f16} explains the behavior of escape velocity of the
particle moving around the black hole. The
shaded region in Fig. \ref{f16} corresponds to escape velocity of the particle and
the solid curved line represents the minimum velocity required to
escape from the vicinity of the black hole and the unshaded region
is for bound motion around the black hole. In Fig. \ref{a16} we
have plotted the escape velocity for different values of energy
$\mathcal{E}$. Fig. \ref{a16} shows that the possibility for the
particle having greater energy has more possibility to escape from the vicinity of black hole as compared to the particle with lesser
value of energy. Escape velocity for different values of $c_{1}$ is
plotted in Fig. \ref{b16}, it shows that greater
the value of $c$ greater will be the escape velocity of the particle,
provided that the energy of the orbiting particle after collision is
less then $U_{max}$ otherwise it would be captured by the black
hole. One can conclude that the presence of dark energy might play a
crucial role in the transfer mechanism of energy to the particle
during its motion in the ISCO. In Fig. \ref{c16} a comparison
of the escape velocities for different values of magnetic field ,$b$, is done. It
can be seen that greater the strength of
magnetic field the possibility for a particle to escape increases. It
can be concluded that the key role in the transfer mechanism  of
energy to the particle for escape from the vicinity of black hole is
played by the magnetic field  which is present in the accretion disc. This is in agreement with  the result of \cite{1,2}.


\section{Summary and Conclusion}
\begin{itemize}
\item
We have studied the dynamics of a neutral and a charged particle in
the vicinity of Schwarzschild black hole surrounded by quintessence
matter. It is known that the quintessence is the candidate for dark
energy and the black hole metric which we have studied was derived
by Kiselev \cite{Kiselev}.
\item
We  have studied the motion of a neutral  particle in the absence of
magnetic field and the dynamics of a charged particle in the
presence of magnetic field in the vicinity of black hole in detail.
\item
We have discussed the energy conditions for the stable circular
orbits and for unstable circular orbits around the black hole.
\item The ISCO for a massive neutral particle around a Schwarzschild-like
black hole occurs at $r=4M$.
 \item We have found that ISCO shifts
closer to the event horizon due to presence of dark energy and
magnetic field as compared to Schwarzschild black hole. This is the
indication that the force due to dark energy is attractive which is
in agreement with the results of \cite{q2}.
\item
Center of mass energy expressions are derived for the colliding
particles near horizons. It is found that CME is finite for the
particle colliding in the vicinity of Kiselev solution.
\item
We have derived the formula for escape velocity of the particles, after collision.
\item
 The equations of motion have been solved numerically and we have plotted the radial velocity of the particle.
\item
The Lyapunov exponent, which gives the instability
time scale for the geodesics of the particle, is calculated. Therefore we have
concluded that the instability of the circular orbits around
Schwarzschild black hole is more compared to Kiselev solution, in the presence of
magnetic field.
\item
We have derived the effective force acting on the particle due to
dark energy and mentioned the conditions under which the force is attractive or repulsive.
\end{itemize}

\subsection*{Acknowledgments} The authors would like to thank the
reviewer and Prof. Dr. Azad Akhter Siddiqui for fruitful comments to improve
this work.

\pagebreak



\end{document}